\documentclass[journal=fmp]{CUP-JNL-DTM}

\usepackage{graphicx}
\usepackage{multicol,multirow}
\usepackage{amsmath,amssymb,amsfonts}
\usepackage{mathrsfs}
\usepackage{amsthm}
\usepackage{rotating}
\usepackage{appendix}
\usepackage{ifpdf}
\usepackage[T1]{fontenc}
\usepackage{newtxtext}
\usepackage{newtxmath}
\usepackage{textcomp}
\usepackage{xcolor}
\usepackage[colorlinks,allcolors=blue]{hyperref}
\usepackage[english=british]{csquotes}

\usepackage{xspace}
\usepackage{booktabs}
\usepackage{tabularx}
\usepackage{array}
\usepackage{makecell}
\usepackage{colortbl}
\usepackage{pifont}
\usepackage{siunitx}
\usepackage{pdflscape}

\usepackage{tikz}
\usepackage{pgfplots}
\usepackage{pgfgantt}
\usepackage{forest}
\usetikzlibrary{shapes, arrows.meta, positioning, calc, decorations.pathreplacing, patterns, shadows, fadings, backgrounds}
\pgfplotsset{compat=1.18}

\usepackage[caption=false]{subfig}

\usepackage{caption}        
\usepackage{placeins}       
\usepackage[htt]{hyphenat}  
\usepackage{enumitem}       

\usepackage[export]{adjustbox}

\usepackage[framemethod=tikz]{mdframed}

\graphicspath{{images/}}

\theoremstyle{definition}

\usepackage[acronym,toc]{glossaries}
\makeglossaries
\newacronym{ca}{CA}{cellular automaton}
\newacronym{gol}{GoL}{Game of Life}
\newacronym{ot}{OT}{outer-totalistic}
\newacronym{vn}{vN}{von Neumann}

\newcommand{\ac}[1]{\gls{#1}}

\newcommand{\acp}[1]{\glspl{#1}}

\jname{J. R. Soc. Interface}
\articletype{Research Article}
\jyear{2026}

\begin{document}


\begin{Frontmatter}

\title{The Self-Replication Phase Diagram: Mapping Where Life Becomes Possible in Cellular Automata Rule Space}

\author[1]{Don Yin\thanks{ORCID: \href{https://orcid.org/0000-0002-8971-1057}{0000-0002-8971-1057}}}

\authormark{Yin}

\address[1]{\orgdiv{School of Clinical Medicine}, \orgname{University of Cambridge}, \orgaddress{\city{Cambridge}, \country{United Kingdom}}}


\abstract{What substrate features allow life? We exhaustively classify all 262,144 outer-totalistic binary cellular automata rules with Moore neighbourhood for self-replication and produce phase diagrams in the $(\lambda, F)$ plane, where $\lambda$ is Langton's rule density and $F$ is a background-stability parameter. Of these rules, 20,152 (7.69\%) support pattern proliferation, concentrated at low rule density ($\lambda \approx 0.15$\textendash$0.25$) and low-to-moderate background stability ($F \approx 0.2$\textendash$0.3$), in the weakly supercritical regime (Derrida coefficient $\mu = 1.81$ for replicators vs.\ $1.39$ for non-replicators). Self-replicating rules are more approximately mass-conserving (mass-balance 0.21 vs.\ 0.34), and this generalises to $k{=}3$ Moore rules. A three-tier detection hierarchy (pattern proliferation, extended-length confirmation, and causal perturbation) yields an estimated 1.56\% causal self-replication rate. Self-replication rate increases monotonically with neighbourhood size under equalised detection: von Neumann 4.79\%, Moore 7.69\%, extended Moore 16.69\%. These results identify background stability and approximate mass conservation as the primary axes of the self-replication phase boundary.}

\end{Frontmatter}




\section{Introduction}\label{ch:intro}

What substrate features allow life? Self-replication, the ability of a pattern to produce copies of itself, is a fundamental signature of living systems. In \acp{ca}, some rule configurations support self-replicating structures while others do not \citep{vonneumann1966, langton1984}. Despite decades of study, no work has systematically mapped where in substrate parameter space self-replication becomes possible.

Langton \citep{langton1990} established that \ac{ca} dynamics undergo a phase transition as a function of rule density $\lambda$, with complex behaviour that concentrates at an intermediate critical value $\lambda_c$ \citep[see also][]{li1990}. Wootters and Langton \citep{wootters1990} showed that this transition sharpens as the number of cell states $k$ increases. However, $\lambda$ alone is a noisy predictor of individual rule behaviour \citep{mitchell1993}. Sakai et al.\ \citep{sakai2004} introduced a second parameter $F$ that measures quiescent-background stability and showed, for one-dimensional $k{=}4$ totalistic rules, that the order\textendash chaos boundary occupies only $\sim$11\% of the $F$ range at fixed $\lambda$, a tenfold sharpening relative to $\lambda$ alone.

Self-replicating \acp{ca} have been studied in specific substrates: Langton \citep{langton1984} designed 8-state loops, Byl \citep{byl1989} reduced this to 6 states, Reggia et al.\ \citep{reggia1993} observed emergent replication in \ac{ca} space, Chou and Reggia \citep{chou1997} showed that self-replicating structures can emerge from random initial conditions, and Yang \citep{yang2024} and Hintze and Bohm \citep{hintze2025} discovered spontaneous distributed self-replication in a binary \ac{ca}. Brown and Sneppen \citep{brown2025} recently catalogued replicators in a 3-state \ac{gol} extension with varying survival thresholds. In continuous systems, Papadopoulos et al.\ \citep{papadopoulos2024} found self-replication near phase boundaries in multi-channel Lenia, and Plantec et al.\ \citep{plantec2025} showed that mass conservation increases the frequency of self-maintaining patterns.

Complementary efforts have characterised the full Life-like rule space ($2^{18} = 262{,}144$ outer-totalistic binary rules with Moore neighbourhood) without targeting self-replication. Eppstein \citep{eppstein2010} classified the non-B0 subset (roughly half) of these rules by growth and decay properties. Turney \citep{turney2020} computed behavioural feature vectors for all 262,144 rules, and Kumar et al.\ \citep{kumar2024} used CLIP-based open-endedness scores to search Life-like rule space via optimisation in their ASAL framework. None of these studies mapped the prevalence of self-replication across rule-space parameters.

Yet no study has parameterised the space of substrates themselves (varying cell states, neighbourhood geometry, rule density, and background stability simultaneously) and mapped where self-replication occurs. Here we present a systematic mapping. We exhaustively classify all 262,144 outer-totalistic binary \ac{ca} rules with Moore neighbourhood for self-replication and produce phase diagrams in the $(\lambda, F)$ plane. We find that:

\begin{enumerate}[leftmargin=*]
    \item Self-replication concentrates at low rule density ($\lambda \approx 0.15$\textendash$0.25$) and low-to-moderate background stability ($F \approx 0.2$\textendash$0.3$), in the weakly supercritical regime; Derrida coefficient analysis yields $\mu = 1.81$ for replicators versus $1.39$ for non-replicators, which places self-replication above the edge of chaos ($\mu = 1$) rather than below it. Under an equalised detection protocol, self-replication rate increases monotonically with neighbourhood size (\ac{vn} 4.79\% [3.6\textendash 6.3\%], Moore 7.69\%, extended Moore 16.69\% [15.97\textendash 17.44\%]).
    \item 20,152 of 262,144 Life-like rules (7.69\%) support pattern proliferation.
    \item Rules with self-replication are more approximately mass-conserving than those without (mass-balance score 0.21 vs.\ 0.34), even without an explicit conservation constraint.
    \item A three-tier detection hierarchy finds that 97.8\% of rules with pattern proliferation are confirmed under an extended-length rescreen, and 20.8\% of those contain causally fragile self-replicating patterns, for an estimated 1.56\% causal self-replication rate. Rules with self-replication show stronger spatial synergy (O-information $\Omega = -0.30$ vs.\ $-0.24$, $p < 10^{-12}$), though multivariate analysis shows this signal is largely accounted for by mass-balance (logistic regression AUC $= 0.85$, mass-balance dominant). Approximate mass conservation generalises to $k{=}3$ Moore rules ($d = -1.04$, $p = 4.2 \times 10^{-227}$).
\end{enumerate}

\section{Methods}\label{ch:methods}

\subsection{Substrate parameterisation}

We study two-dimensional \acp{ca} on a square lattice with periodic boundary conditions. Each cell takes one of $k$ states and is updated synchronously according to an \ac{ot} rule: the new state depends only on the current centre-cell state and the sum of neighbour states.

We sweep four parameters:

\begin{enumerate}[leftmargin=*]
    \item \textbf{Cell states} $k \in \{2, 3\}$: controls information capacity per cell. We enumerate $k=2$ exhaustively and sample $k=3$.
    \item \textbf{Neighbourhood geometry}: \ac{vn} ($|N|=5$) and Moore ($|N|=9$).
    \item \textbf{Rule density} $\lambda$: the fraction of non-quiescent entries in the rule table \citep{langton1990}. We sweep $\lambda$ from 0.05 to $1 - 1/k$ in 20 steps.
    \item \textbf{Background stability} $F$: adapted from the background-stability concept of Sakai et al.\ \citep{sakai2004}, re-weighted for outer-totalistic rules. $F_{\text{OT}}$ quantifies how aggressively the rule destroys quiescent-background regions. For \ac{ot} rules:
    \[
    F_{\text{OT}} = \frac{\sum_{s=0}^{S_{\max}} w(s) \cdot \mathbb{1}[T(0, s) \neq 0]}{\sum_{s=0}^{S_{\max}} w(s)}, \quad w(s) = \left(1 - \frac{s}{S_{\max}}\right)^{2}
    \]
    where $T(0, s)$ is the rule-table output for centre state 0 with neighbour sum $s$, $S_{\max}$ is the maximum possible sum, and $w(s)$ weights low-neighbour-sum configurations more heavily (a quiescent centre surrounded by mostly quiescent neighbours is the canonical ``background'' condition).
\end{enumerate}

For $k=2$ with Moore neighbourhood, the \ac{ot} rule space contains $2^{18} = 262{,}144$ rules (the ``Life-like'' rules), which we enumerate exhaustively. For $k=2$ with \ac{vn} neighbourhood, the space contains $2^{10} = 1{,}024$ rules, also enumerated exhaustively. For $k=3$ Moore, extended Moore, and C4 rule samples, we drew 10,000 rules each, stratified by $\lambda$ (500 rules per $\lambda$ bin across 20 bins, rejection sampling was used to achieve the target density).

The $(\lambda, F)$ space is a discrete combinatorial lattice, not a continuum. $\lambda = n / (k \cdot |N|)$ takes $k \cdot |N| + 1$ evenly spaced values (19 for $k{=}2$ Moore), with rule counts per level given by $\binom{k \cdot |N|}{n}$. $F$ is a weighted subset sum over the quiescent-centre row; for $k{=}2$ Moore it takes 147 distinct values. The lattice therefore has structurally empty regions (no rule can exist there), visible as white cells in the phase diagrams.

\subsection{Defining self-replication: three tiers}

The phase boundary location depends on the definition of self-replication. Motivated by the formal distinction between pattern copying and self-replication in Cotler et al.\ \citep{cotler2025}, we operationalise three detection tiers:

\textbf{Tier 1 (pattern proliferation):} A rule supports pattern proliferation if a bounded non-quiescent pattern $P$ appears in strictly increasing copy count (at least 3 increases) over $T_{\max}$ timesteps. Detection uses connected-component labelling with rotation- and reflection-invariant canonical hashing.

\textbf{Tier 2 (extended-length confirmation):} The same pattern-proliferation criterion is applied at extended simulation length (512 steps, 8 checkpoints instead of 4) and provides an independent confirmation of detection robustness.

\textbf{Tier 3 (causal self-replication):} A replicating pattern is isolated on an empty grid, confirmed to self-replicate alone, then each cell is individually deleted and the simulation re-run. If $\geq$50\% of single-cell deletions prevent replication ($N=10$ trials), the pattern is causally fragile; its structure is necessary for replication, not merely sufficient.

\begin{figure}[htbp]
    \centering
    \includegraphics[width=\textwidth]{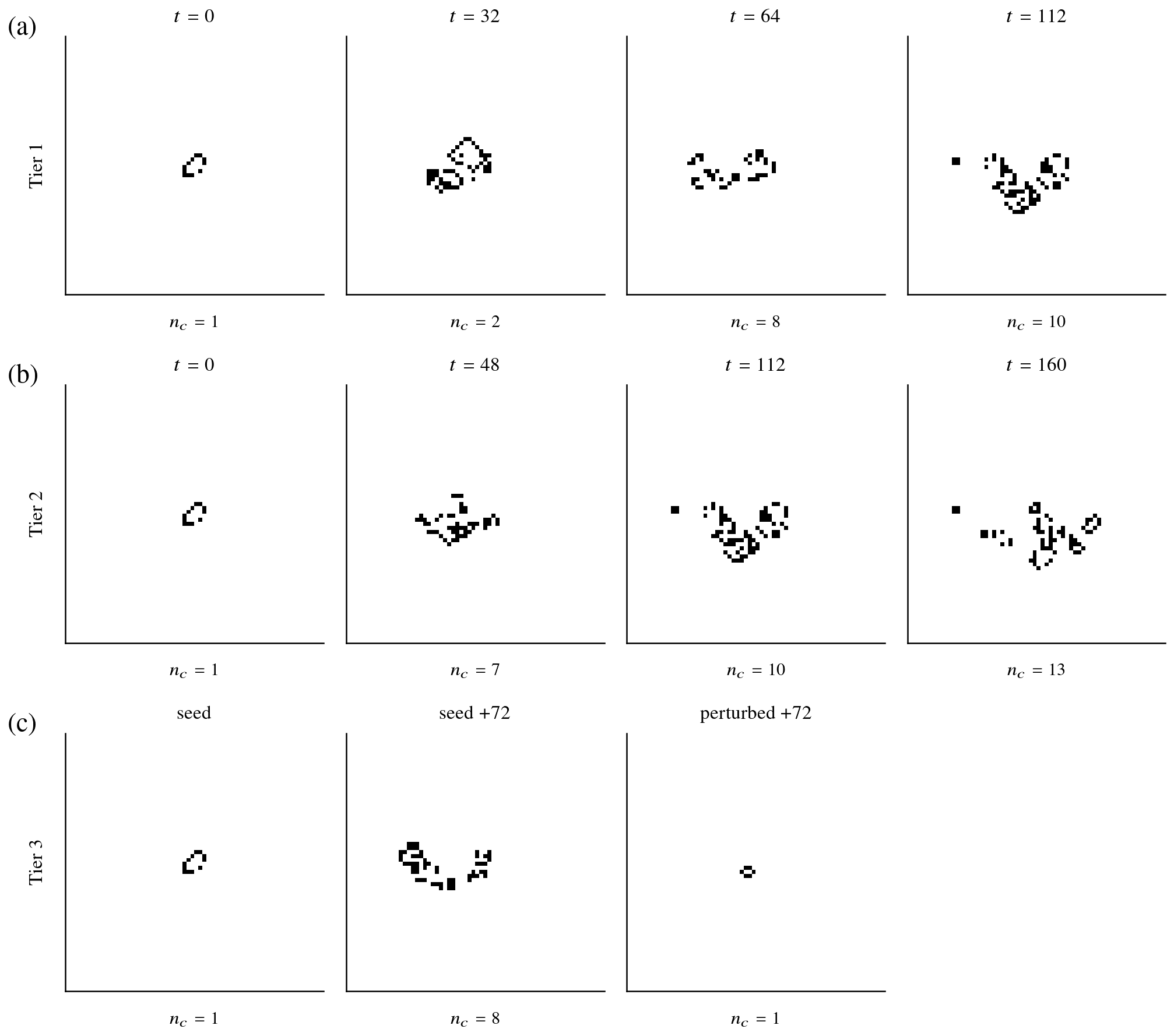}
    \caption{Worked example of the three-tier detection hierarchy using HighLife (B36/S23). \textit{(a)}~Tier~1: component count $n_c$ increases from 1 to 10 over 112 steps, which triggers the pattern-proliferation criterion. \textit{(b)}~Tier~2: the same pattern tested at extended length (160 steps); proliferation continues ($n_c = 13$), which confirms sustained replication. \textit{(c)}~Tier~3: a seed pattern is isolated on an empty grid and self-replicates ($n_c = 1 \to 8$); a single-cell deletion from the seed prevents replication ($n_c = 1$), which establishes causal fragility.}
    \label{fig:tier-examples}
\end{figure}

Figure~\ref{fig:tier-examples} illustrates the three tiers using the HighLife (B36/S23) replicator as a worked example.

\subsection{Detection pipeline}

\textbf{Stage 1 (screening):} Each rule is tested from two initial densities ($D_0 = 0.15$ and $0.35$) on a $64 \times 64$ grid for 256 timesteps with snapshots every 64 steps. A rule is flagged if \emph{any} indicator triggers: (A) monotonically increasing component count over 3+ checkpoints, or (B) any multi-cell canonical component hash accumulating 3+ instances across all snapshots.

\textbf{Stage 2 (extended-length rescreen):} For flagged rules, the same proliferation criterion is re-applied at 512 steps with 8 checkpoints (twice the initial screen length) and provides independent confirmation of detection robustness.

\textbf{Stage 3 (causal test):} For rules confirmed at extended length, perturbation experiments destroy parent patterns early (when the grid is sparse) via minimal single-cell deletions.

\subsection{Derived measures}

At each rule, we compute:

\begin{itemize}[leftmargin=*]
    \item \textbf{Mass-balance score}: the mean absolute mass change per step, normalised by grid area:
    \[
    M = \frac{1}{T} \sum_{t=1}^{T} \frac{|m(t) - m(t-1)|}{W \times H}
    \]
    where $m(t) = \sum_{i,j} c_{ij}(t)$ is the total live-cell count at step $t$, and $W \times H = 64 \times 64$ is the grid size. We compute $M$ from a dedicated 64-step simulation at initial density $D_0 = 0.15$. Lower values indicate more approximately conservative dynamics.
    \item \textbf{O-information} $\Omega = \text{TC} - \text{DTC}$ \citep{rosas2019}: computed from the joint distribution of $3 \times 3$ spatial patches across simulation snapshots. Negative $\Omega$ indicates synergy dominance (the whole carries more information than the sum of its parts); positive $\Omega$ indicates redundancy.
    \item \textbf{Spatial entropy}: mean local Shannon entropy computed over $3 \times 3$ patches across the grid.
\end{itemize}

\subsection{Derrida coefficient}

To locate the edge of chaos experimentally, we computed the Derrida coefficient $\mu$ \citep{derrida1986} for sampled rules. For each rule, we initialised a $32 \times 32$ grid at 50\% density, then ran 100 single-cell perturbation trials: in each trial, one randomly chosen cell was flipped, and both the original and perturbed grids were evolved for $T = 10$ steps with snapshots at every step. The Hamming distance $\delta_t$ between the two trajectories was recorded at each step. The Derrida coefficient is defined as the mean ratio of consecutive distances:
\[
\mu = \langle \delta_{t+1} / \delta_t \rangle
\]
averaged over all time steps where $\delta_t > 0$ and over all perturbation trials. $\mu < 1$ indicates ordered (subcritical) dynamics, $\mu = 1$ marks the critical point, and $\mu > 1$ indicates chaotic (supercritical) dynamics.

\subsection{Computational details}

Simulations use FFT-based convolution for neighbour-sum computation and achieve 118\,ms per rule including detection. The full $k=2$ Moore census (262,144 rules) completed in 8.7 hours on a single workstation.

The detection pipeline was tested against known benchmarks: the \ac{gol} (B3/S23) and HighLife (B36/S23) are correctly identified as positive for self-replication, while both a dead rule (all-zero table) and a high-$\lambda$ chaotic rule are correctly identified as negative. All code is available at \url{https://github.com/Don-Yin/self-replication}.

\section{Results}\label{ch:results}

\subsection{Exhaustive census of Life-like rules}

We exhaustively tested all 262,144 \ac{ot} binary \ac{ca} rules with Moore neighbourhood for pattern proliferation. Of these, 20,152 rules (7.69\%) support pattern proliferation. For comparison, the \ac{vn} neighbourhood ($|N|=5$, 1,024 rules exhaustively tested) yields a tier-1 rate of 4.79\% under an equalised detection protocol (Section~\ref{sec:vn-moore}).

\subsection{Phase diagram in the $(\lambda, F)$ plane}

\begin{figure}[htbp]
    \centering
    \includegraphics[width=\textwidth]{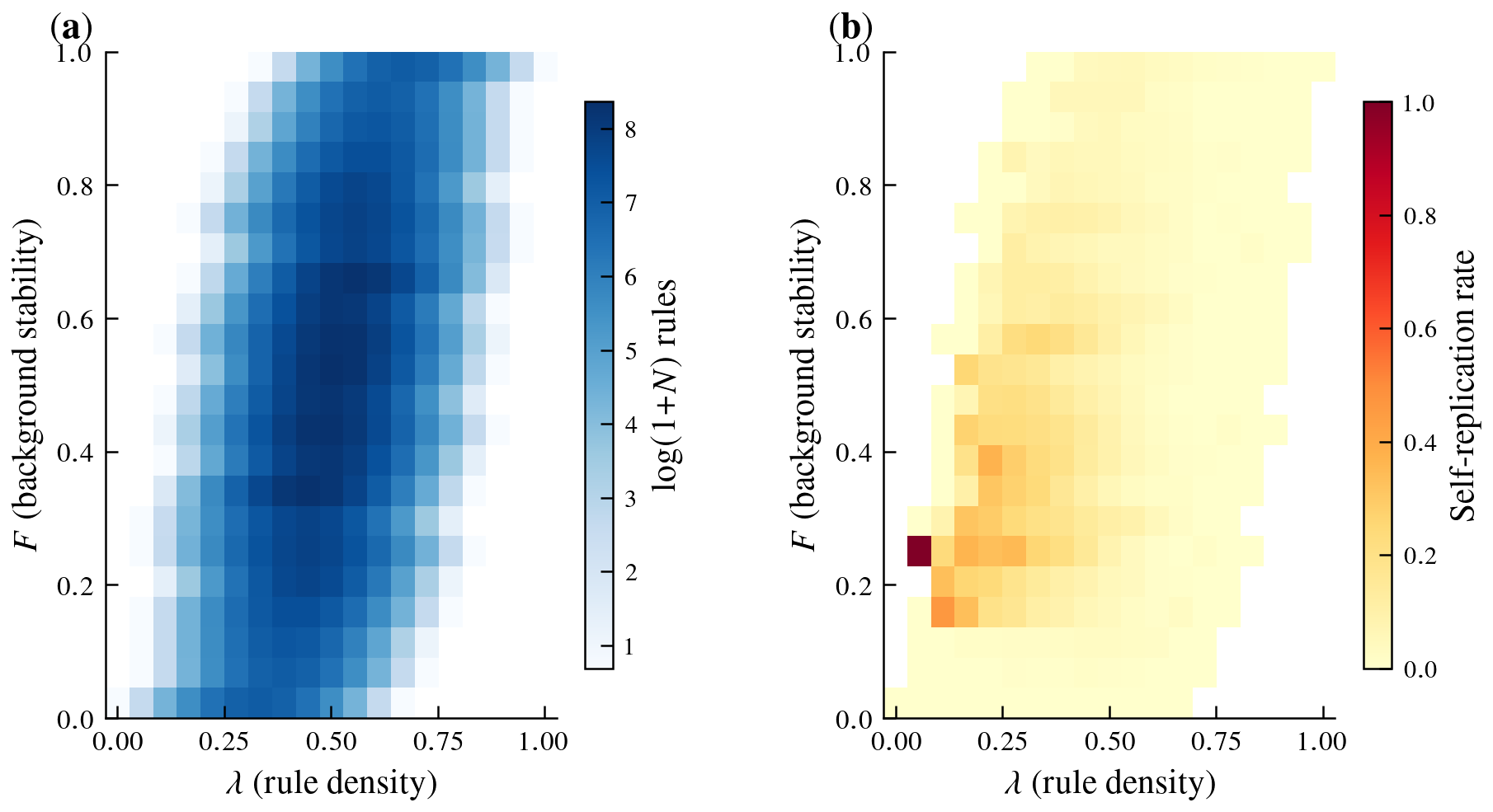}
    \caption{Phase diagram for all 262,144 Life-like (outer-totalistic $k=2$ Moore) rules. Columns correspond to the 19 exact $\lambda = n/18$ values; $F$ is binned into 22 intervals (the 147 distinct $F$ values, which are weighted subset sums of binary rule-table entries, are too dense to resolve individually). White cells indicate structurally empty regions where no rule exists. \textit{(a)}~Rule density (log scale): the binomial concentration at $\lambda \approx 0.5$ reflects $\binom{18}{9} = 48{,}620$ rules. \textit{(b)}~Self-replication rate: the ``island of life'' at low $\lambda$ and low-to-moderate $F$ concentrates in the weakly supercritical regime identified by the Derrida analysis (Section~\ref{sec:derrida}).}
    \label{fig:phase-diagram}
\end{figure}

Figure~\ref{fig:phase-diagram} shows the self-replication phase diagram in the $(\lambda, F)$ plane for all 262,144 Life-like rules. Self-replication concentrates in a localised region at low rule density ($\lambda \approx 0.1$--$0.3$) and low-to-moderate background stability ($F \approx 0.15$--$0.35$).

\begin{figure}[htbp]
    \centering
    \includegraphics[width=0.65\textwidth]{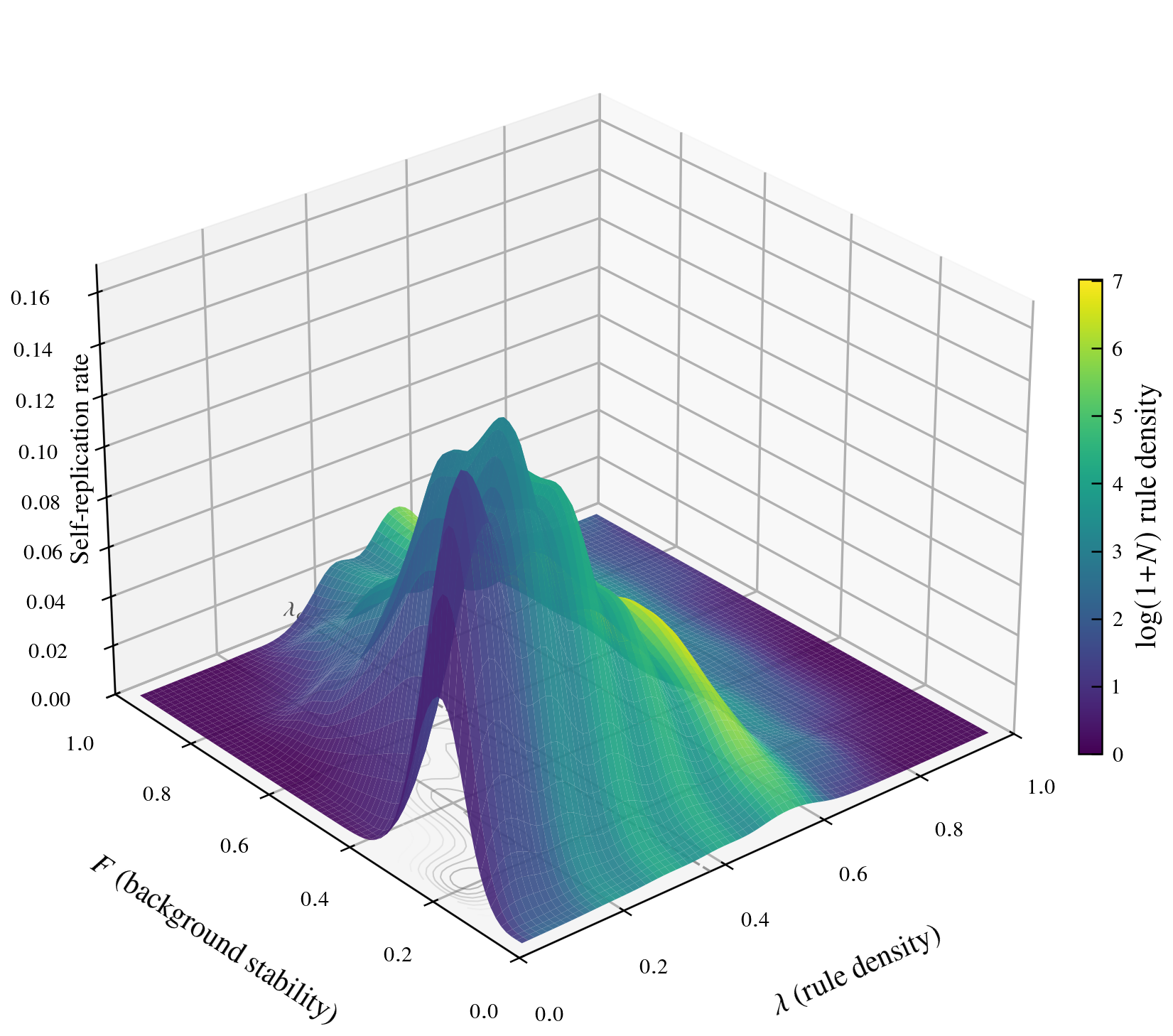}
    \caption{Smoothed 3D surface of the self-replication rate over the $(\lambda, F)$ plane (Gaussian filter $\sigma=0.8$, cubic spline interpolation; colour encodes log rule density). The sharp peak at low $\lambda$ and low $F$ shows that life-supporting rules occupy a narrow island in parameter space. The Derrida analysis (Section~\ref{sec:derrida}) places this peak in the weakly supercritical regime ($\mu \approx 1.4$\textendash$1.8$), just above the edge of chaos. Note: both $\lambda$ and $F$ are discrete; the smooth surface is a visual aid, not a claim of continuity.}
    \label{fig:surface3d}
\end{figure}

Figure~\ref{fig:surface3d} shows a smoothed 3D surface interpolated from the same data, which makes the localised ``island of life'' visually apparent: a sharp peak rises from an otherwise flat landscape of zero replication.

Self-replication peaks at $\lambda \approx 0.15$\textendash$0.25$. The Derrida analysis (Section~\ref{sec:derrida}) shows that the edge of chaos ($\mu = 1$) occurs at $\lambda \approx 0.05$\textendash$0.10$, so self-replicating rules concentrate in the weakly supercritical regime ($\mu \approx 1.4$\textendash$1.8$), just above the critical threshold rather than deep in the chaotic phase. This is consistent with the biological requirement that replicators need substrates with enough dynamical capacity to propagate structure, yet not so much disorder that patterns are destroyed.

The ``island of life'' is qualitatively robust to the choice of $F$-weighting scheme. Under uniform, linear, and quadratic weightings of birth/survival contributions to $F$, the island persists in all three cases with peak replication rates of $\sim$14\%, but the $F$-axis position of the peak shifts (0.225, 0.275, and 0.375 respectively). The island's existence is therefore a structural feature of Life-like rule space, while its precise $F$-coordinate depends on the weighting convention.

\begin{figure}[htbp]
    \centering
    \includegraphics[width=0.7\textwidth]{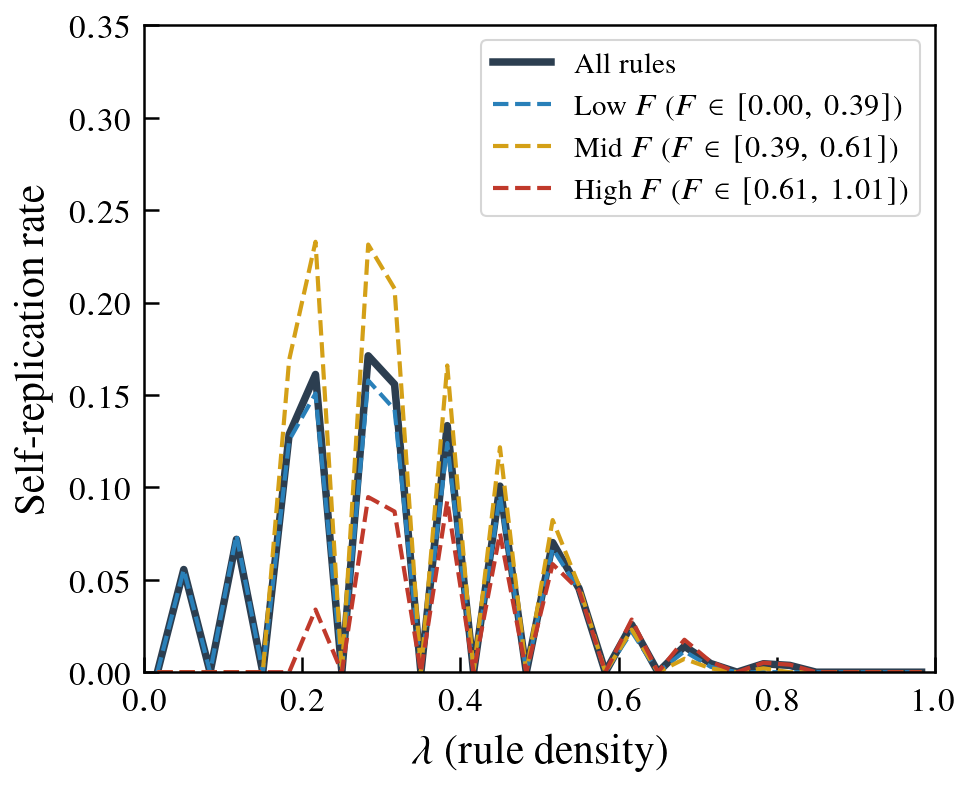}
    \caption{Self-replication rate versus $\lambda$, overall (black) and conditioned on $F$ terciles (coloured). The peak shifts with $F$ level, which indicates that background stability adds discriminatory power beyond $\lambda$ alone.}
    \label{fig:lambda-profiles}
\end{figure}

\begin{figure}[htbp]
    \centering
    \includegraphics[width=0.7\textwidth]{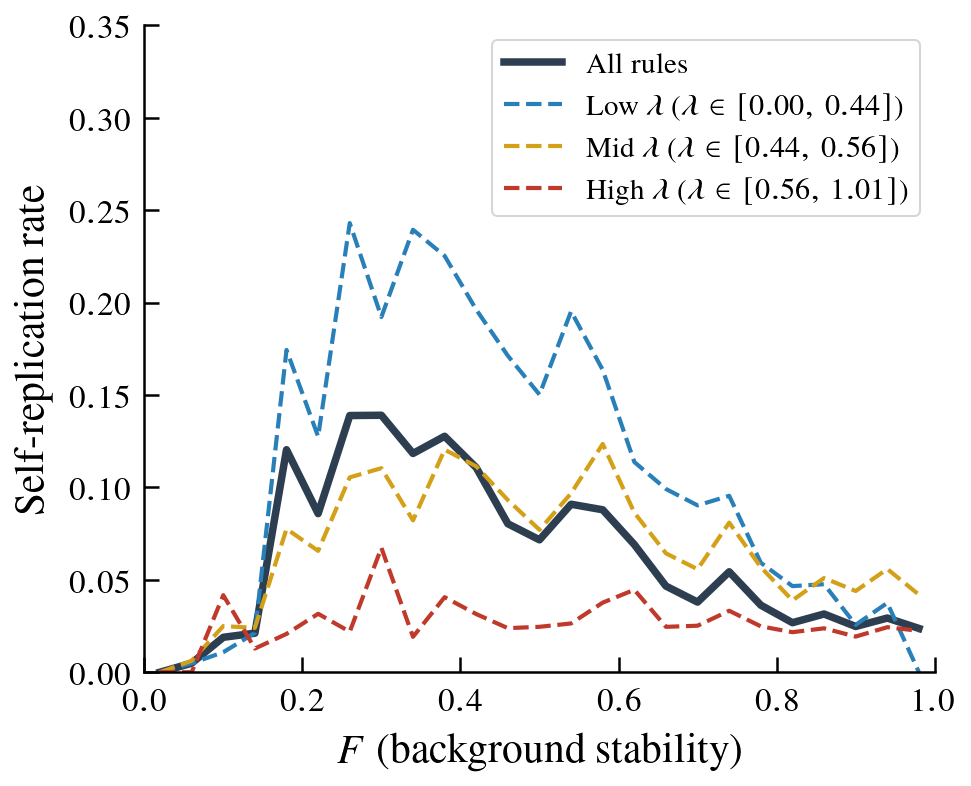}
    \caption{Self-replication rate versus $F$, overall (black) and conditioned on $\lambda$ terciles (coloured). Self-replication peaks at low-to-moderate $F$ ($\approx 0.2$--$0.4$) and drops at both extremes, consistent with the localised island in the phase diagram.}
    \label{fig:f-marginal}
\end{figure}

Figure~\ref{fig:lambda-profiles} shows the self-replication rate as a function of $\lambda$, conditioned on $F$ terciles. Figure~\ref{fig:f-marginal} shows the complementary view: self-replication rate as a function of $F$, conditioned on $\lambda$ terciles. Self-replication peaks at low-to-moderate $F$ ($\approx 0.2$--$0.4$) and drops at both extremes.

\subsection{Derrida coefficient and the edge of chaos}\label{sec:derrida}

To locate the edge of chaos experimentally rather than by heuristic, we computed the Derrida coefficient $\mu$ for each rule by measuring the mean Hamming-distance ratio between successor pairs of initially close configurations. A value $\mu < 1$ indicates ordered (sub-critical) dynamics; $\mu = 1$ marks the critical point; $\mu > 1$ indicates chaotic (supercritical) dynamics.

\begin{figure}[htbp]
    \centering
    \includegraphics[width=\textwidth]{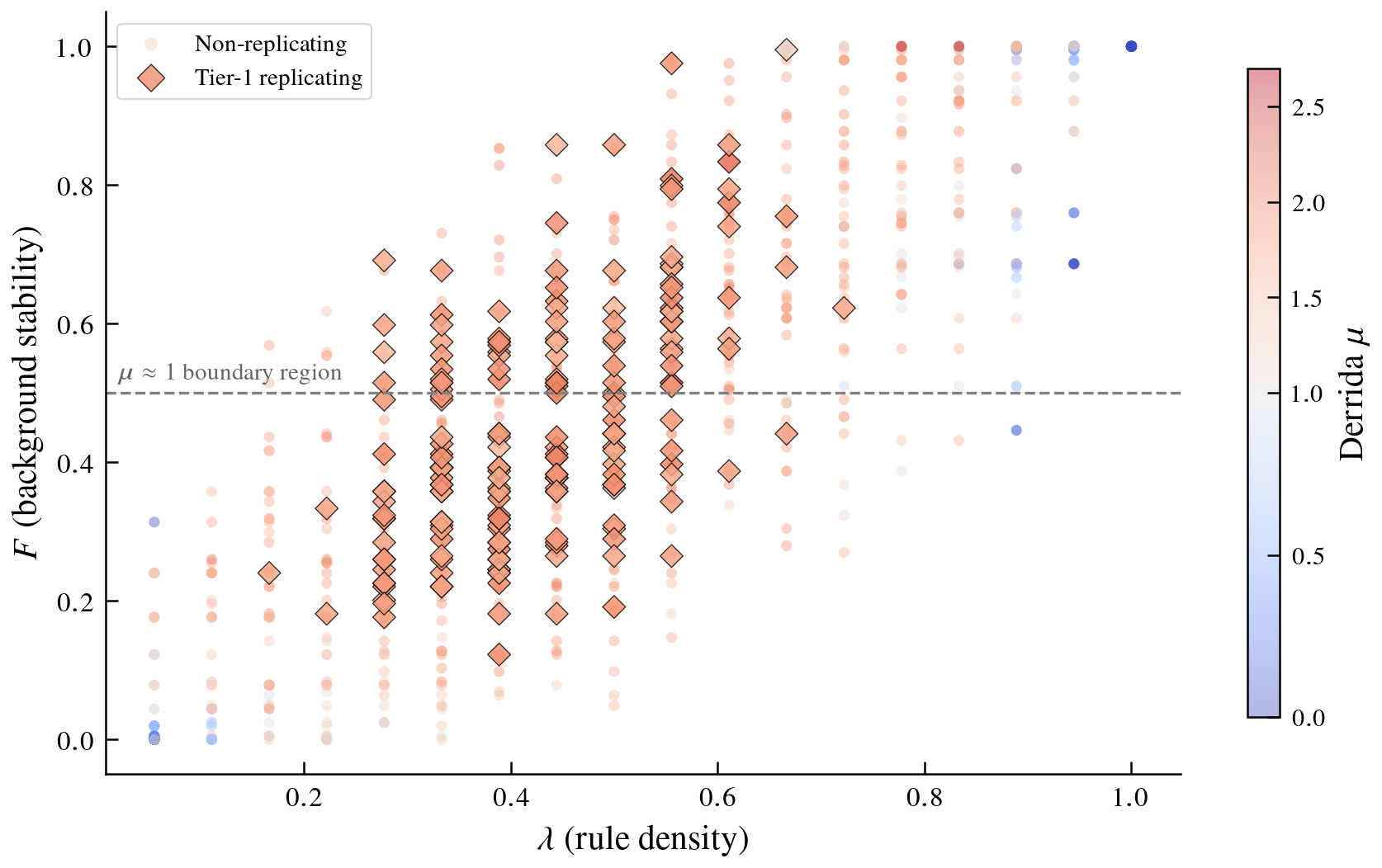}
    \caption{Derrida coefficient $\mu$ in the $(\lambda, F)$ plane. Colour encodes $\mu$ (blue: subcritical, $\mu < 1$; red: supercritical, $\mu > 1$; dashed line: $\mu = 1$ boundary). Tier-1-positive rules (diamonds) cluster in the red region at $\mu \approx 1.4$\textendash$1.8$, above the critical threshold.}
    \label{fig:derrida-phase}
\end{figure}

Tier-1-positive rules have a mean Derrida coefficient $\mu = 1.81$, compared to $\mu = 1.39$ for tier-1-negative rules; both groups are supercritical on average, but replicators are more so. Binning rules by $\lambda$ shows a monotonic profile: $\mu$ rises from 0.76 ($\lambda < 0.1$) to 1.41 ($0.1$\textendash$0.2$), 1.63 ($0.2$\textendash$0.3$), 1.78 ($0.3$\textendash$0.4$), and 1.82 ($0.4$\textendash$0.5$). The edge of chaos ($\mu = 1$) falls at $\lambda \approx 0.05$\textendash$0.10$, which places the self-replication peak ($\lambda \approx 0.15$\textendash$0.25$) firmly in the weakly supercritical regime. Self-replicating rules are not anomalously sub-critical; they sit just above the critical threshold, where dynamics are expansive enough to propagate copies yet structured enough to maintain pattern coherence. Figure~\ref{fig:derrida-phase} shows the Derrida coefficient across the $(\lambda, F)$ plane.

\subsection{Von Neumann versus Moore neighbourhood}\label{sec:vn-moore}

Under the original detection protocol (4 initial conditions per density, 512 steps), the \ac{vn} neighbourhood ($|N|=5$) yields a tier-1 rate of 33.2\% (340/1,024), higher than Moore (7.69\%). However, the original protocol allocated more detection effort to \ac{vn} rules than to Moore rules. To control for this, we re-ran the full \ac{vn} census under an equalised protocol matched to the Moore screening conditions (1 initial condition per density, 256 steps). Under equalised detection, the \ac{vn} tier-1 rate drops to 4.79\% (95\% CI\footnote{All reported CIs are Wilson score intervals.}: [3.6\%, 6.3\%]; 49/1,024).

\begin{figure}[htbp]
    \centering
    \includegraphics[width=\textwidth]{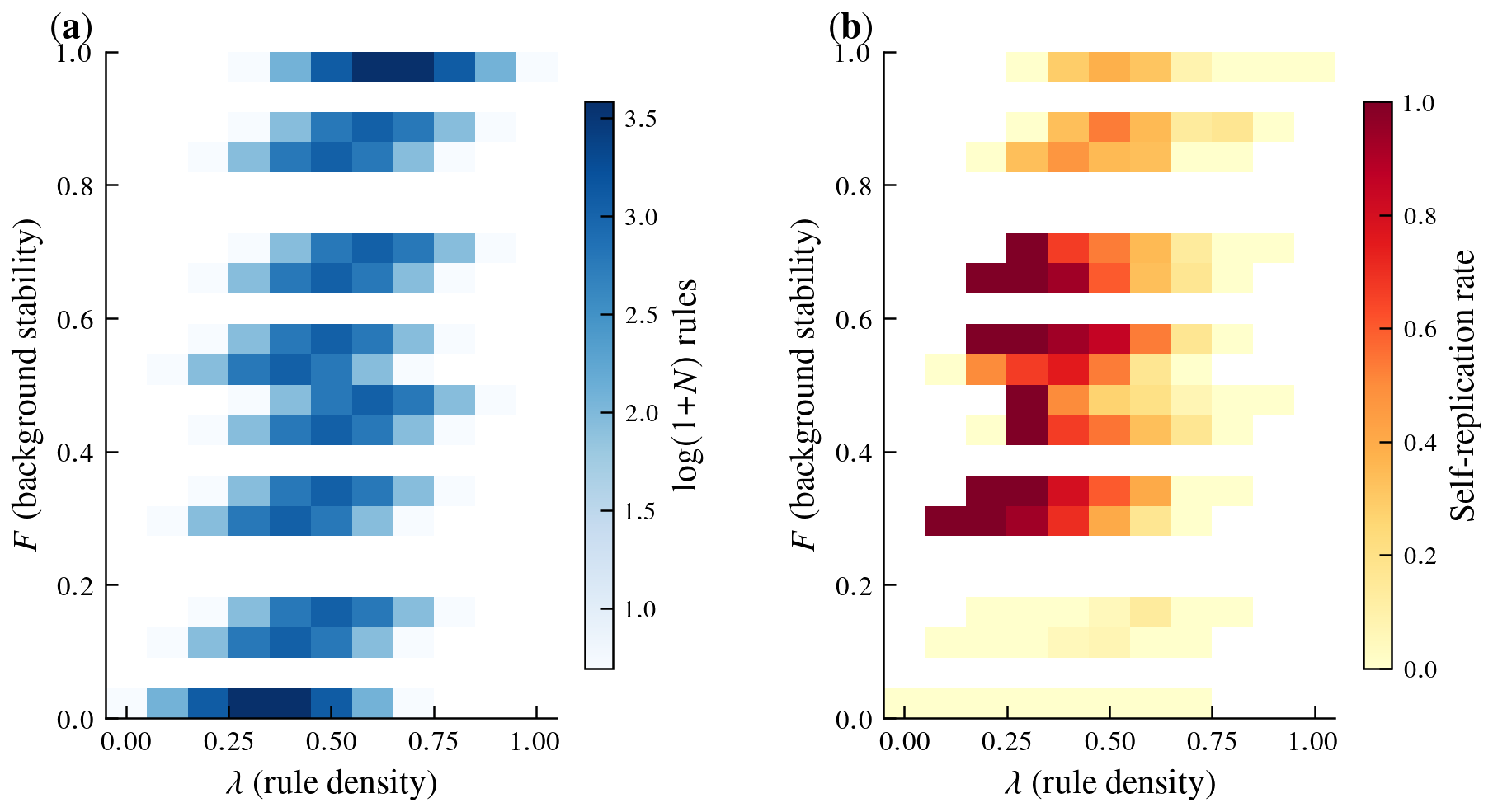}
    \caption{Phase diagram for all 1,024 outer-totalistic $k=2$ \ac{vn} rules on the discrete $(\lambda, F)$ lattice. Under equalised detection (1~IC/density, 256 steps), the tier-1 rate is 4.79\%. The lattice is coarser ($\lambda = n/10$, 11 values) because the \ac{vn} rule table has only 10 entries.}
    \label{fig:vn-phase}
\end{figure}

We additionally sampled 10,000 $k{=}2$ extended Moore rules ($|N|=25$). The self-replication rate is 16.69\% (95\% CI: [15.97\%, 17.44\%]). Under equalised detection, the relationship between neighbourhood size and self-replication is monotonically increasing with $|N|$: \ac{vn} 4.79\% [3.6\textendash 6.3\%] ($|N|=5$) $<$ Moore 7.69\% ($|N|=9$) $<$ extended Moore 16.69\% [15.97\textendash 17.44\%] ($|N|=25$). Larger neighbourhoods provide each cell with a wider receptive field, which enables richer local computations and increases the fraction of rule space supporting self-replication. The higher \ac{vn} rate under the original protocol reflects greater detection sensitivity (more initial conditions, longer simulation), not necessarily false positives; the equalised protocol sacrifices sensitivity for cross-substrate comparability.

\subsection{Derived measures at the boundary}

\begin{figure}[htbp]
    \centering
    \includegraphics[width=\textwidth]{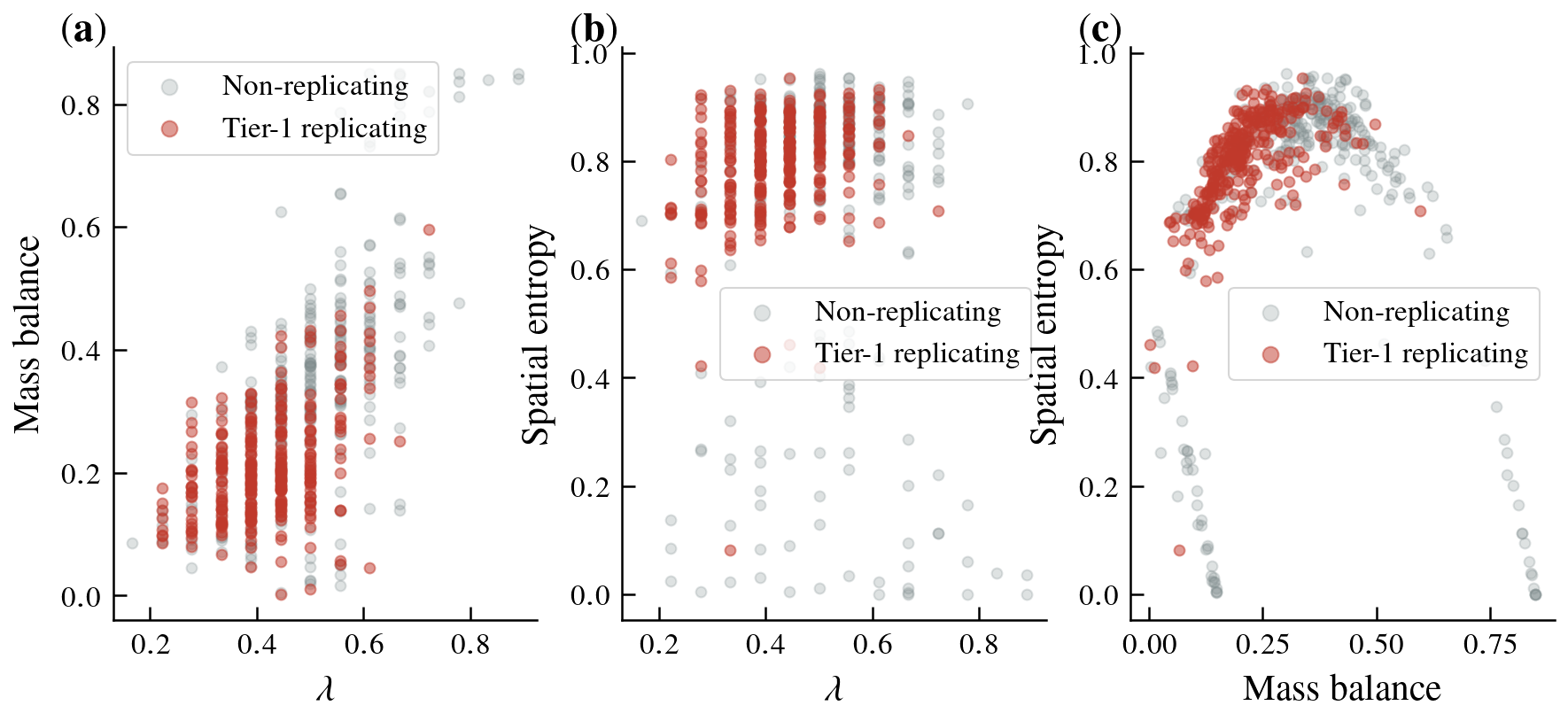}
    \caption{Scatter plots comparing replication-positive (red) and replication-negative (grey) rules. \textit{(a)}~Mass balance versus $\lambda$: replicating rules cluster at lower mass balance (more approximately conserving). \textit{(b)}~Spatial entropy versus $\lambda$. \textit{(c)}~Mass balance versus spatial entropy: the two measures jointly separate the classes.}
    \label{fig:boundary-scatter}
\end{figure}

We sampled 300 replication-positive and 300 replication-negative rules from the Moore census and computed derived measures. Table~\ref{tab:boundary} summarises the comparison and Figure~\ref{fig:boundary-scatter} shows the distributions.

\begin{table}[htbp]
    \centering
    \begin{tabular}{lccccc}
        \toprule
        Measure & Replication+ & Replication$-$ & Difference & $p$ (Mann--Whitney) & Cohen's $d$ \\
        \midrule
        Mass balance & 0.210 & 0.345 & $-0.135$ & $1.9 \times 10^{-26}$ & $-0.94$ \\
        O-information ($\Omega$) & $-0.299$ & $-0.236$ & $-0.063$ & $9.3 \times 10^{-13}$ & $-0.27$ \\
        Dual total correlation & 0.774 & 0.621 & $+0.153$ & $1.2 \times 10^{-7}$ & $+0.27$ \\
        Total correlation & 0.474 & 0.385 & $+0.090$ & $6.7 \times 10^{-4}$ & $+0.17$ \\
        Spatial entropy & 0.798 & 0.719 & $+0.080$ & $0.055^{\dagger}$ & $+0.39$ \\
        \bottomrule
    \end{tabular}
    \caption{Derived measures for rules at the self-replication phase boundary (300 rules per class, Mann--Whitney $U$ test). Mass balance is the strongest discriminator ($d = -0.94$, large effect); O-information shows a smaller univariate effect ($d = -0.27$) that does not survive multivariate control (Section~\ref{sec:multivariate}). $^{\dagger}$Not significant at $\alpha = 0.05$.}
    \label{tab:boundary}
\end{table}

The mass-balance difference is the strongest signal: self-replicating rules are 1.6$\times$ more approximately mass-conserving than non-replicating rules, \emph{despite no explicit conservation constraint}. This is consistent with the hypothesis that conservation facilitates self-replication \citep{plantec2025, papadopoulos2025}, even in discrete rule systems where exact number conservation forces intrinsically one-dimensional dynamics for binary adjacent-cell rules \citep{wolnik2019}, which severely limits the structural repertoire available.

To control for potential confounding between self-replication status and location in $(\lambda, F)$ space, we repeated the comparison with $\lambda$/$F$-matched negative controls (nearest-neighbour matching in the $(\lambda, F)$ plane). The mass-balance signal survived matching (Cohen's $d = -0.94$, $p = 1.9 \times 10^{-26}$), as did O-information ($d = -0.27$, $p = 9.3 \times 10^{-13}$). A permutation test (10,000 shuffles) yielded $p < 10^{-4}$ for mass-balance and $p = 0.001$ for O-information.

\subsection{O-information: synergy at the boundary}

We computed O-information \citep{rosas2019} for 300 replication-positive and 300 replication-negative rules. We estimated total correlation (TC), dual total correlation (DTC), and their difference $\Omega = \text{TC} - \text{DTC}$ from the joint distribution of $3 \times 3$ spatial patches across simulation snapshots. Negative $\Omega$ indicates synergy dominance (the whole carries more information than the sum of its parts); positive $\Omega$ indicates redundancy.

\begin{figure}[htbp]
    \centering
    \includegraphics[width=\textwidth]{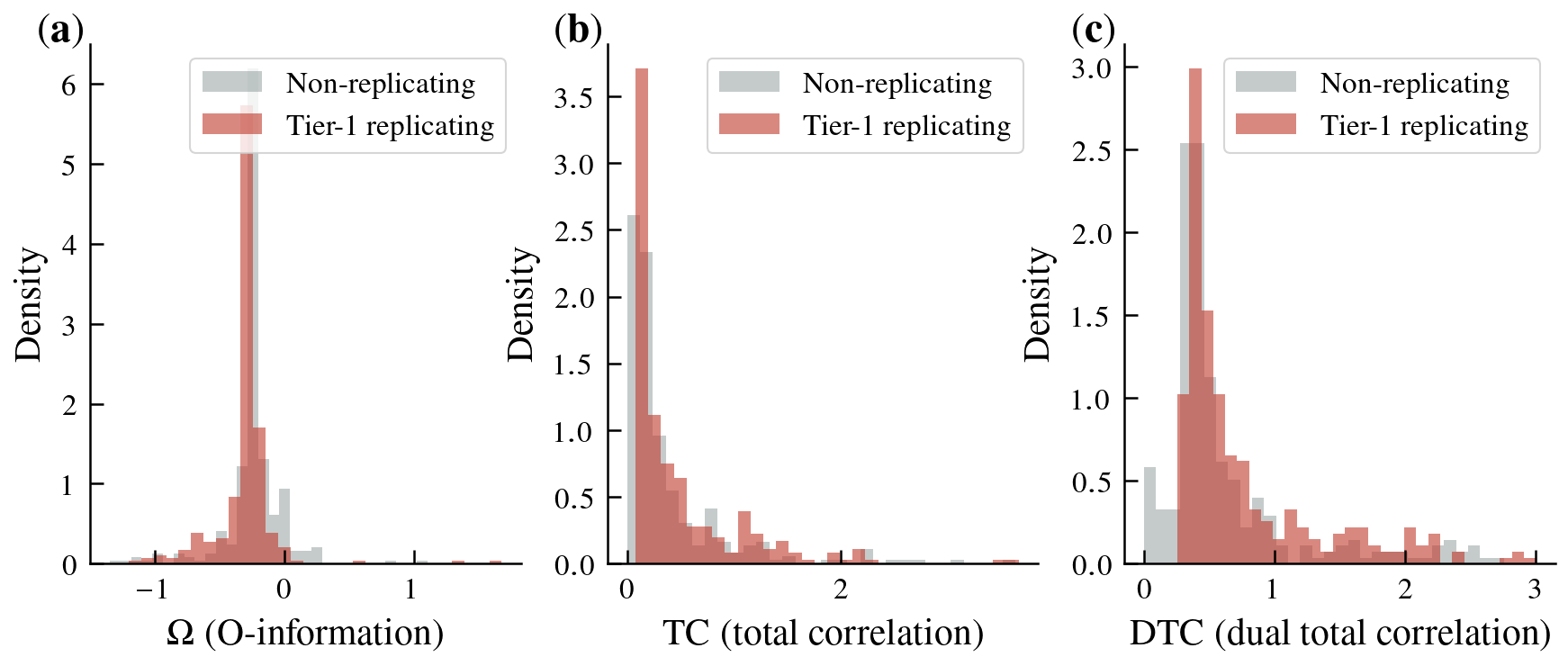}
    \caption{Information-theoretic measures at the self-replication boundary. Left: O-information (negative = synergy). Centre: total correlation (TC). Right: dual total correlation (DTC). Replicating rules (red) show more synergy (more negative $\Omega$) and heavier tails in both TC and DTC.}
    \label{fig:oinfo}
\end{figure}

Both groups are synergy-dominated ($\Omega < 0$), but self-replicating rules show \emph{more} synergy: $\Omega = -0.30$ for replication-positive versus $-0.24$ for replication-negative rules. The dual total correlation is $\sim$25\% higher in replicating rules (DTC $= 0.77$ vs.\ $0.62$), which indicates stronger higher-order statistical dependencies. Figure~\ref{fig:oinfo} shows the distributions.

The synergy signal suggests that self-replicating substrates exhibit information structure that cannot be decomposed into pairwise correlations; the spatial pattern of a replicator carries higher-order dependencies. However, as the multivariate analysis in Section~\ref{sec:multivariate} shows, this signal does not add predictive power beyond mass-balance, which indicates that O-information and approximate conservation share substantial variance.

\subsection{Multi-state extension: $k = 3$}

\begin{figure}[htbp]
    \centering
    \includegraphics[width=0.8\textwidth]{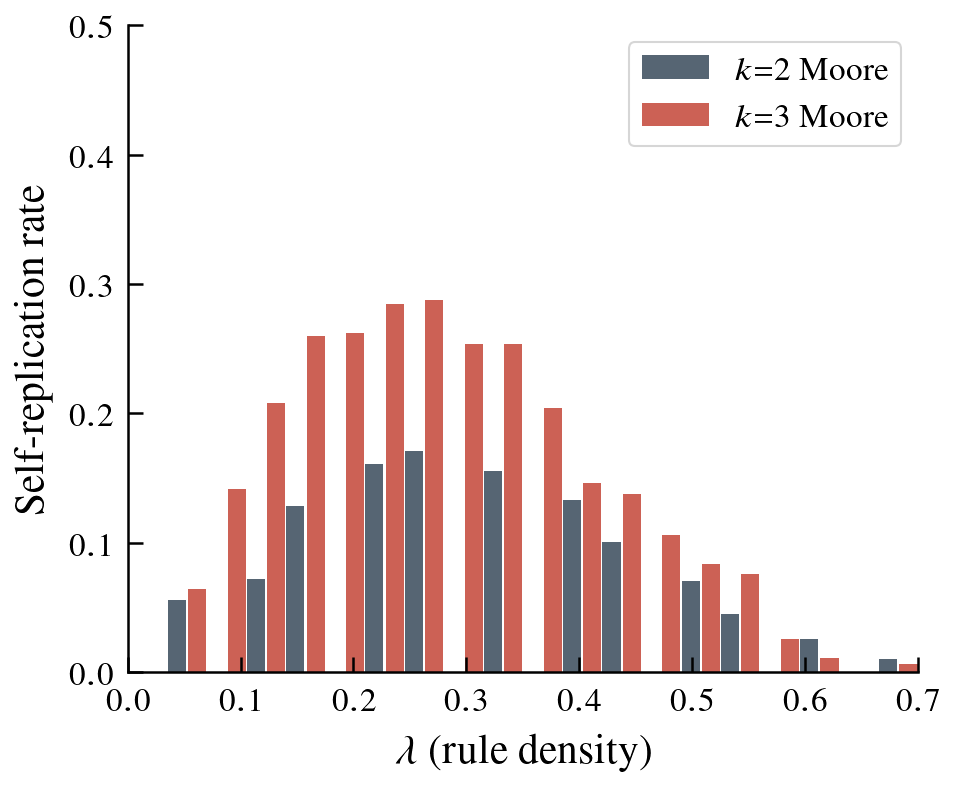}
    \caption{Self-replication rate versus $\lambda$ for $k=2$ (blue, exhaustive) and $k=3$ (red, sampled). The $k=3$ rate is consistently higher and peaks at a slightly higher $\lambda$, consistent with a richer state alphabet that enables more compact replicators.}
    \label{fig:k2-vs-k3}
\end{figure}

To test whether the phase boundary shifts with state count, we sampled 10,000 \ac{ot} $k=3$ Moore rules across 20 $\lambda$ bins. The self-replication rate is 15.55\% (95\% CI: [14.85\%, 16.28\%]), approximately double the $k=2$ rate (7.69\%). Figure~\ref{fig:k2-vs-k3} compares the $\lambda$ profiles.

The doubling from $k=2$ to $k=3$ indicates that state count is an axis of the self-replication phase diagram: more states shift the boundary outward and increase the fraction of rule space that supports self-replication.

\subsection{Rule symmetry: totalistic constraint helps}\label{sec:c4}

To test whether the outer-totalistic constraint restricts or facilitates self-replication, we sampled 10,000 rotationally-symmetric (C4) $k=2$ Moore rules from the $2^{140}$-dimensional space where the Outlier \ac{ca} \citep{yang2024} resides. C4 rules preserve 4-fold rotational symmetry but break reflection symmetry and are not constrained to depend only on neighbour sums.

The C4 self-replication rate is 3.57\% (95\% CI: [3.22\%, 3.95\%]; 357 / 10,000), \emph{lower} than the \ac{ot} rate of 7.69\%. This is counterintuitive: relaxing the totalistic constraint, which nominally allows richer dynamics, reduces the frequency of self-replication. One interpretation is that the totalistic constraint acts as a regulariser. By restricting the rule table to neighbour counts (18 entries) rather than specific neighbour configurations (512 entries), it suppresses chaotic dynamics and increases the fraction of rules in the structured regime where self-replication is possible. As with the conservation finding, structural constraints that reduce dynamical freedom can paradoxically \emph{increase} the prevalence of complex behaviour.

\subsection{Tier-2 confirmation: extended-length rescreen}\label{sec:tier2}

To distinguish genuine self-replication from transient artefacts, we rescreened 1,000 replication-positive rules at extended simulation length (512 steps, 8 checkpoints; Section~\ref{ch:methods}). Of these, 978 (97.8\%, 95\% CI: [96.7\%, 98.6\%]) were confirmed as positive under the extended-length screen.

This high confirmation rate indicates that the vast majority of rules flagged by the initial screen produce sustained proliferation at longer time horizons, not merely transient artefacts or boundary-condition copying. The 2.2\% non-confirmation rate provides an upper bound on the screening error.

\subsection{Tier-3 causal test: pattern-level perturbation}\label{sec:tier3}

We tested all 978 tier-2-confirmed rules for causal self-replication. For each rule, we (1) identified a replicating component from an early simulation snapshot, (2) placed it in isolation on an empty grid and confirmed it self-replicates alone, and (3) systematically deleted single cells from the seed pattern (10 trials per rule) and checked whether replication was prevented.

Of the 978 rules tested, 203 (20.8\%, 95\% CI: [18.3\%, 23.4\%]) passed the causal self-replication test (prevention rate $\geq 0.5$). The remaining rules fell into three categories: 571 (58.4\%) were \emph{robust replicators} whose seed patterns replicate in isolation but tolerate single-cell perturbations; 119 (12.2\%) exhibited \emph{distributed replication}, where patterns proliferate from random initial conditions but not from isolated seed patterns; and 85 (8.7\%) could not be tested due to density or extraction limitations.

The distributed-replication class (12.2\%) contains rules that support pattern proliferation only through cooperative multi-component dynamics, not from any single self-replicating structure. This is consistent with the distributed selfhood observed by Hintze and Bohm \citep{hintze2025} in the Outlier cellular automaton, and suggests that distributed self-replication may be more common than previously assumed in Life-like rule space.

Combining the pass rates across the three-tier detection hierarchy (7.69\% at tier~1, 97.8\% [96.7\textendash 98.6\%] at tier~2, 20.8\% [18.3\textendash 23.4\%] at tier~3) yields an estimated \textbf{1.56\%} (95\% CI: [1.2\%, 2.0\%]) of all 262,144 Life-like rules as causal self-replicators.

\subsection{Detection robustness}\label{sec:sensitivity}

To test whether the census results depend on detection hyperparameters, we re-ran tier-1 detection on 5,000 rules (2,500 original positives, 2,500 negatives) at 512 steps with varied snapshot intervals (32, 64, 128) and minimum-increase thresholds (2, 3, 4, 5). At the census settings (interval 64, threshold 3), 99.9\% of original tier-1 rules are recovered. At the stricter setting (interval 64, threshold 4), 96.3\% are still recovered while the false-positive rate among original negatives drops from 66\% to 40\%. The separation between original positives and negatives is consistent across all 12 parameter combinations, which supports the conclusion that the phase-diagram structure is robust even though the absolute tier-1 rate shifts with detection stringency.

\subsection{Multivariate prediction of self-replication}\label{sec:multivariate}

A logistic regression trained on ($\lambda$, $F$, mass-balance, spatial entropy, O-information) achieves AUC $= 0.85$ (5-fold cross-validation) for predicting self-replication status. Mass-balance is the dominant predictor (standardised coefficient $-2.09$), followed by spatial entropy ($+1.35$) and $\lambda$ ($-0.42$). O-information contributes negligibly after controlling for the other features (coefficient $-0.01$), which suggests that the O-information signal in Table~\ref{tab:boundary} is largely accounted for by shared variance with mass-balance and $\lambda$.

\begin{figure}[htbp]
    \centering
    \includegraphics[width=0.5\textwidth]{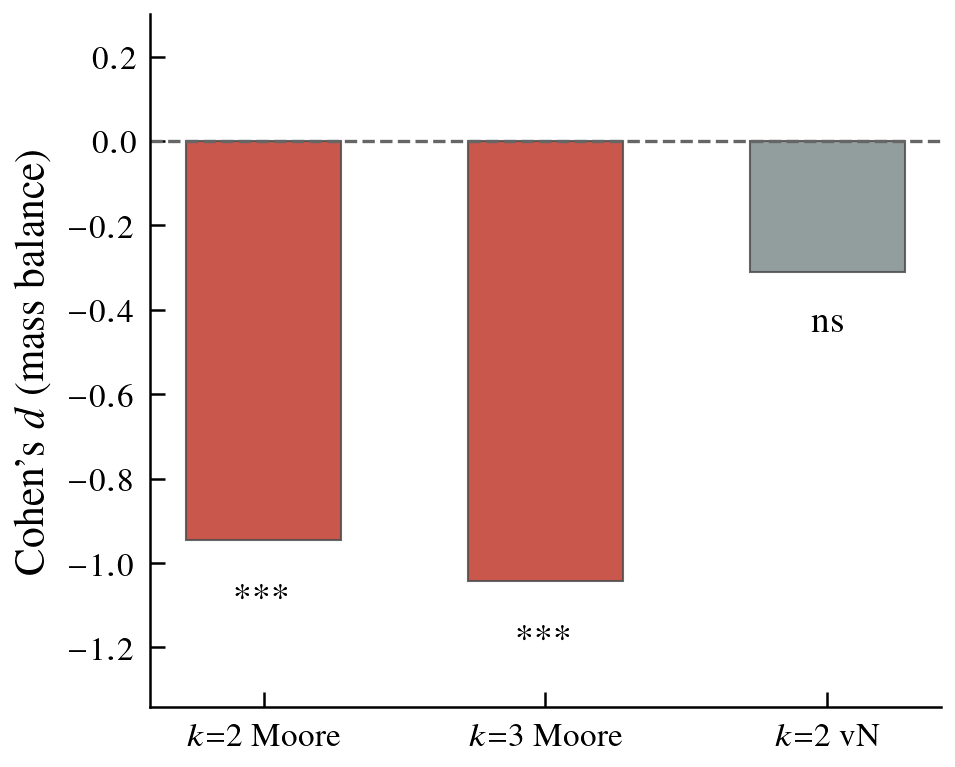}
    \caption{Cohen's $d$ for the mass-balance difference between replication-positive and replication-negative rules across three substrates. $k{=}2$ Moore and $k{=}3$ Moore show large, significant effects; $k{=}2$ von Neumann is weaker and non-significant.}
    \label{fig:cross-substrate}
\end{figure}

The mass-balance signal generalises across substrates. For $k=3$ Moore rules, the replication-positive/negative mass-balance difference is large (Cohen's $d = -1.04$, $p = 4.2 \times 10^{-227}$), exceeding the $k=2$ effect size. For \ac{vn} rules (with tier-1 labels from the original protocol, $n_{\text{pos}} = 340$), the effect is weaker and non-significant ($d = -0.31$, $p = 0.30$), likely reflecting both the smaller rule space (1,024 rules) and the reduced statistical power of the \ac{vn} parameter space. Approximate mass conservation is therefore a cross-substrate predictor of self-replication, at least for rule spaces with sufficient combinatorial richness. Figure~\ref{fig:cross-substrate} summarises the cross-substrate comparison.

\subsection{Cross-comparison with ASAL open-endedness}\label{sec:asal}

To test whether self-replication correlates with visual open-endedness, we computed the Spearman rank correlation between each rule's tier-1 replication status and its ASAL open-endedness score \citep{kumar2024}, which quantifies the diversity of CLIP-embedded visual patterns produced across simulation trajectories. The correlation is effectively zero ($\rho = -0.002$, $p = 0.24$). Self-replication is therefore independent of CLIP-based visual open-endedness: rules that produce diverse visual patterns are no more (or less) likely to support self-replication than rules that converge to uniform or repetitive fields. This dissociation suggests that the computational requirements for self-replication (approximate conservation, moderate dynamical complexity) are orthogonal to those for visual diversity as measured by foundation-model embeddings.

\subsection{Verification at larger scale}

To test whether the pattern-proliferation results are finite-size artefacts, we re-ran the 100 lowest-$\lambda$ replication-positive rules on a $256 \times 256$ grid for 8,192 timesteps (compared to $64 \times 64$ and 256 steps in the census). All 100 rules (100\%) were positive at the larger scale, with no false positives. This indicates that the patterns detected in the census represent scale-robust proliferation dynamics.

\subsection{Example replicators}

\begin{figure}[htbp]
    \centering
    \includegraphics[width=\textwidth]{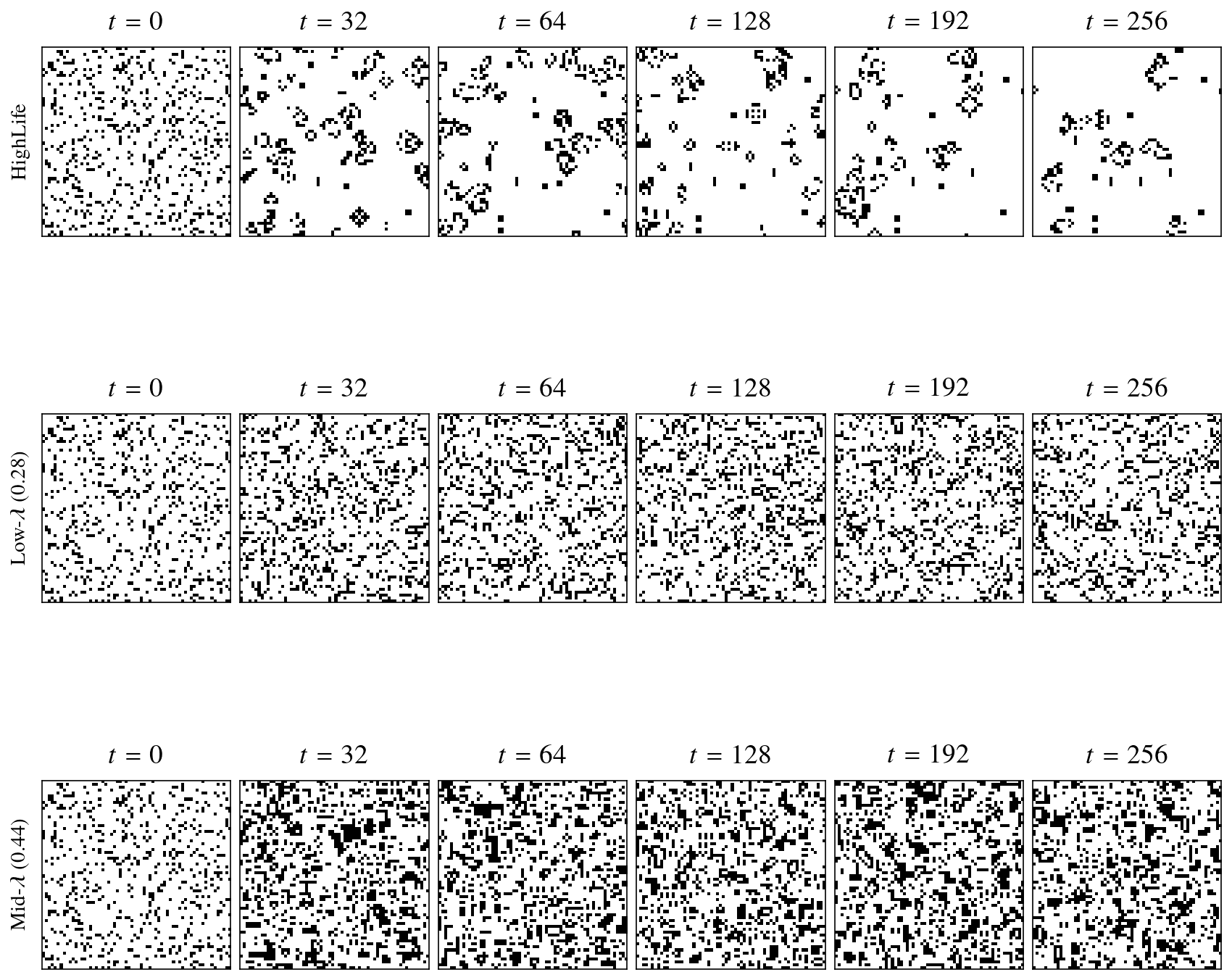}
    \caption{Time-lapse visualisations of three self-replicators on $64 \times 64$ grids from density $D_0 = 0.15$. Columns show snapshots at $t = 0, 32, 64, 128, 192, 256$. Top: HighLife (B36/S23). Middle: low-$\lambda$ replicator. Bottom: mid-$\lambda$ replicator. All exhibit localised structures that proliferate over time.}
    \label{fig:replicators}
\end{figure}

Figure~\ref{fig:replicators} shows time-lapse visualisations of three example self-replicators: HighLife (B36/S23), a low-$\lambda$ replicator ($\lambda = 0.28$), and a mid-$\lambda$ replicator ($\lambda = 0.44$). All three show the characteristic pattern of localised structures that spawn copies into quiescent space.

\section{Discussion}\label{ch:discussion}

\subsection{Beyond the edge of chaos: conservation as the binding constraint}

The conventional expectation, following Langton \citep{langton1990}, was that self-replication should concentrate at the edge of chaos, where dynamical activity is maximal yet still controllable. Our data refine this picture. Self-replication peaks at $\lambda \approx 0.15$\textendash$0.25$, where the Derrida coefficient is already supercritical: $\mu = 1.81$ for tier-1-positive rules versus $\mu = 1.39$ for tier-1-negative rules at matched $\lambda$. The edge of chaos ($\mu = 1$) itself falls at $\lambda \approx 0.05$\textendash$0.10$, \emph{below} the self-replication peak.

This means self-replicators are not poised at criticality; they inhabit a weakly supercritical regime. Moreover, replicating rules are \emph{more} dynamically active than non-replicating rules at the same $\lambda$, not less. The Derrida gap ($\Delta \mu \approx 0.4$) is consistent with a picture in which replication requires enough perturbation growth to propagate spatial copies, while rules that merely conserve structure without amplifying it cannot sustain proliferation.

If dynamical activity alone were sufficient, the self-replication rate should peak near the Derrida maximum, well above $\lambda = 0.25$. It does not, because a second constraint binds first: approximate mass conservation. Rules at higher $\lambda$ are supercritical but non-conserving; their dynamics destroy patterns faster than copies can form. The binding constraint is therefore conservation, not criticality. Self-replication requires weak supercriticality \emph{plus} approximate conservation, and the narrow $\lambda$ window where both conditions are jointly satisfied defines the island of life in the phase diagram.

This reframes the Langton hypothesis. Dynamical activity at or above the edge of chaos is necessary (rules in the ordered regime ($\mu < 1$) almost never self-replicate), but it is not sufficient. Conservation provides the additional structural constraint that channels supercritical dynamics into pattern-preserving replication rather than pattern-destroying chaos.

\subsection{Background stability as a hidden variable}

The adapted $F$ parameter (Section~\ref{ch:methods}; cf.\ \citep{sakai2004}) captures a dimension of rule structure that $\lambda$ misses: which entries in the rule table are non-quiescent. Two rules with identical $\lambda$ but different $F$ have the same activity level but differ in whether that activity disrupts quiescent regions. Our phase diagrams show that self-replication is localised in the $(\lambda, F)$ plane; background stability is a relevant axis for characterising life-supporting substrates.

\subsection{Approximate conservation without design}

The finding that self-replicating rules are more approximately mass-conserving connects to recent work on mass-conserving \acp{ca}. Plantec et al.\ \citep{plantec2025} showed that adding exact mass conservation to Lenia transforms the system from one where self-maintaining patterns are rare to one where they are abundant. Papadopoulos and Guichard \citep{papadopoulos2025} reported a similar pattern in continuous \acp{ca} using their MaCE framework. Our data extend this pattern to discrete systems: even without explicit conservation constraints, the rules that happen to approximately conserve mass are those most likely to support self-replication. This association survives $\lambda$/$F$ matching (Cohen's $d = -0.94$) and permutation testing ($p < 10^{-4}$).

The conservation effect generalises beyond binary Moore rules. For $k{=}3$ outer-totalistic Moore rules, the mass-balance effect is even stronger ($d = -1.04$, $p = 4.2 \times 10^{-227}$), which suggests that conservation becomes a more powerful discriminator as the state alphabet grows. For the \ac{vn} neighbourhood, the effect is weaker ($d = -0.31$), though this estimate draws on only 49 tier-1-positive rules under the equalised protocol and may be underpowered. Across these three substrates, approximate conservation is a convergent property of life-supporting rules, not merely an engineering convenience imposed by design.

\subsection{Mass-balance as the primary discriminator}

A logistic regression combining $\lambda$, $F$, mass-balance, spatial entropy, and O-information achieves AUC $= 0.85$ for predicting self-replication status, with mass-balance as the dominant feature (standardised coefficient $-2.09$). O-information, despite a significant univariate signal ($d = -0.27$, $p < 10^{-12}$), contributes negligibly once mass-balance is controlled (coefficient $\approx 0$). This indicates that the spatial synergy observed in replicating rules shares substantial variance with approximate conservation: rules that approximately conserve mass tend also to produce synergy-dominated spatial patterns. Conservation, rather than higher-order information structure per se, appears to be the more fundamental substrate property associated with self-replication.

\subsection{Neighbourhood size: monotonically increasing self-replication}

Under an equalised detection protocol (matching the number of initial conditions and simulation length across neighbourhoods), self-replication rate increases monotonically with neighbourhood size: \ac{vn} ($|N|{=}5$, 4.79\%) $<$ Moore ($|N|{=}9$, 7.69\%) $<$ extended Moore ($|N|{=}25$, 16.69\%). The original \ac{vn} rate of 33.2\% reflects the higher detection sensitivity of a protocol with $4{\times}$ more initial conditions and $2{\times}$ longer simulation time; once these factors are equalised, the \ac{vn} neighbourhood produces the lowest replication rate, not the highest.

The monotonic relationship admits a straightforward interpretation: larger receptive fields provide each cell with more spatial information for encoding replication mechanisms. A $5 \times 5$ extended Moore kernel can represent spatial templates that are inaccessible to the $3 \times 3$ Moore kernel, which in turn can encode richer neighbour interactions than the 4-cell \ac{vn} kernel. The relationship between neighbourhood size and self-replication is thus governed by information capacity, not by a tension between locality and dilution.

A caveat is warranted: the extended Moore rate (16.69\%) derives from a 10,000-rule sample of the $2^{50}$ rule space, whereas the Moore census is exhaustive. Sampling variance may affect the extended Moore estimate, and the true population rate could differ.

\subsection{Self-replication versus open-endedness}

Comparing our exhaustive self-replication census against the ASAL open-endedness scores of Kumar et al.\ \citep{kumar2024} for the same 262,144 rules yields a near-zero correlation ($\rho = -0.002$). CLIP-based visual novelty does not predict self-replication, and self-replicating rules are no more or less ``interesting'' in the ASAL sense than non-replicating rules.

This null result has a clear implication: self-replication is a substrate-structural phenomenon, governed by conservation and mild supercriticality, not by the kind of visual complexity that foundation-model embeddings capture. A rule can produce visually novel, open-ended dynamics without supporting self-replication, and conversely, many self-replicating rules produce visually repetitive patterns that score low on ASAL novelty. The mechanisms that underlie self-replication (approximate conservation, low $\lambda$, appropriate $F$) are structural features of the rule table, not emergent aesthetic properties of the dynamics.

\subsection{Comparison with Brown and Sneppen (2025)}

Brown and Sneppen \citep{brown2025} studied replicators in a three-state extension of the Game of Life (``3-Life'') with varying survival thresholds, and found that self-replication emerges preferentially in low-activity regimes. Their work, selected as a PRE Editors' Suggestion, provides a deep characterisation of replicator dynamics within a single rule family. Our study takes a complementary approach: rather than varying thresholds within one family, we exhaustively survey the full 262,144 outer-totalistic binary rule space and sample 10,000 $k{=}3$ rules to map the self-replication phase boundary across substrate parameters. The convergent finding that replication favours low-activity regimes in both studies strengthens confidence that this is a generic property of Life-like substrates, not an artefact of any particular rule family.

Our $k{=}3$ sweep (15.55\% tier-1 rate, 95\% CI: [14.85\%, 16.28\%]) provides a quantitative complement to the Brown and Sneppen analysis: by sampling across $k{=}3$ outer-totalistic rule space rather than within a single rule, we can estimate the prevalence of self-replication as a function of $\lambda$ and $F$. The mass-balance finding ($d = -1.04$ for $k{=}3$) suggests that approximate conservation may also be relevant in 3-Life, though Brown and Sneppen \citep{brown2025} did not test this directly. The difference between the two studies is one of scope versus depth: we provide a systematic phase diagram across substrate parameters and identify conservation and mild supercriticality as the binding constraints, whereas Brown and Sneppen \citep{brown2025} provide detailed mechanistic analysis of replicator dynamics, including replication cycles and spatial organisation, within a single substrate. The two perspectives are complementary and together suggest that the conditions for self-replication are both narrow and universal across Life-like cellular automata.

\subsection{Limitations}

Our pattern-proliferation definition is permissive, but the extended-length rescreen (Section~\ref{sec:tier2}) reproduces 97.8\% of flagged rules, which bounds the non-confirmation rate at 2.2\%. The causal perturbation test (Section~\ref{sec:tier3}) further narrows this to an estimated 1.56\% true causal self-replication rate. Of the $n = 978$ confirmed rules tested, the fraction classified as ``robust replicators'' (those that replicate in isolation but tolerate single-cell deletions) may represent a genuine category of robust self-replication or may require multi-cell perturbation tests to reveal fragility. Additionally, we tested only outer-totalistic rules. All known non-trivial self-replicators at $k > 2$ use non-totalistic rules with directional information flow \citep{langton1984, hintze2025}. The outer-totalistic restriction may structurally exclude self-replication mechanisms at higher $k$, though our C4 rotationally-symmetric sample (Section~\ref{sec:c4}) begins to probe non-totalistic rule spaces.

The $F$ parameter's axis location in the phase diagram depends on the weighting scheme used to compute it: under alternative weightings the island of life shifts along the $F$ axis, though its existence and localisation are preserved across all weightings tested. This means the precise $F$ coordinates we report are convention-dependent, while the qualitative conclusion, that background stability is a relevant second axis, is robust.

The equalised \ac{vn} census shows that neighbourhood comparisons are sensitive to detection protocol. The raw \ac{vn} rate (33.2\%) and the equalised rate (4.79\%) differ by nearly an order of magnitude, which underscores that protocol-matched comparisons are essential for cross-substrate claims.

\subsection{Outlook}

The immediate next steps are: (1) extending the sweep to $k = 4, 6, 8$ with both outer-totalistic and rotationally-symmetric rules; (2) computing causal emergence \citep{hoel2013} at the phase boundary to test whether the macroscale dynamics of self-replicating rules carry more effective information than their microscale; (3) extending the Derrida analysis to continuous \ac{ca} substrates (Lenia, Flow-Lenia) to test whether the weakly supercritical regime generalises beyond discrete systems; and (4) investigating why the mass-balance effect is weaker for \ac{vn} than for Moore and $k{=}3$, whether this reflects a genuine substrate difference or merely the small sample size under equalised detection.




\begin{Backmatter}

\bibliography{main}

\providecommand{\bysame}{\leavevmode\hbox to3em{\hrulefill}\thinspace}
\providecommand{\MR}{\relax\ifhmode\unskip\space\fi MR }
\providecommand{\MRhref}[2]{%
  \href{http://www.ams.org/mathscinet-getitem?mr=#1}{#2}
}
\providecommand{\href}[2]{#2}
\begin{thebibliography}{10}

\bibitem{brown2025}
F.~D. Brown and Kim Sneppen, \emph{Replicators in {Game-of-Life}-like
  automata}, Physical Review E \textbf{111} (2025), 054306.

\bibitem{byl1989}
John Byl, \emph{Self-reproduction in small cellular automata}, Physica D:
  Nonlinear Phenomena \textbf{34} (1989), 295--299.

\bibitem{chou1997}
Hui-Hsien Chou and James~A. Reggia, \emph{Emergence of self-replicating
  structures in a cellular automata space}, Physica D: Nonlinear Phenomena
  \textbf{110} (1997), 252--276.

\bibitem{cotler2025}
Jordan Cotler, Cl\'{e}ment Hongler, and Barbora Hudcova, \emph{Self-replication
  and computational universality}, arXiv preprint arXiv:2510.08342 (2025).

\bibitem{derrida1986}
Bernard Derrida and Yves Pomeau, \emph{Random networks of automata: A simple
  annealed approximation}, Europhysics Letters \textbf{1} (1986), 45--49.

\bibitem{eppstein2010}
David Eppstein, \emph{Growth and decay in life-like cellular automata}, Game of
  Life Cellular Automata (2010), 71--98.

\bibitem{hintze2025}
Arend Hintze and Christopher Bohm, \emph{Rethinking self-replication: Detecting
  distributed selfhood in the outlier cellular automaton}, npj Complexity
  \textbf{3} (2026), 11.

\bibitem{hoel2013}
Erik~P. Hoel, Larissa Albantakis, and Giulio Tononi, \emph{Quantifying causal
  emergence shows that macro can beat micro}, Proceedings of the National
  Academy of Sciences \textbf{110} (2013), 19790--19795.

\bibitem{kumar2024}
Akarsh Kumar, Samuel Earle, et~al., \emph{{ASAL}: Automated search for
  artificial life}, arXiv preprint arXiv:2412.17799 (2024).

\bibitem{langton1984}
Christopher~G. Langton, \emph{Self-reproduction in cellular automata}, Physica
  D: Nonlinear Phenomena \textbf{10} (1984), 135--144.

\bibitem{langton1990}
\bysame, \emph{Computation at the edge of chaos: Phase transitions and emergent
  computation}, Physica D: Nonlinear Phenomena \textbf{42} (1990), 12--37.

\bibitem{li1990}
Wentian Li, Norman~H. Packard, and Christopher~G. Langton, \emph{Transition
  phenomena in cellular automata rule space}, Physica D: Nonlinear Phenomena
  \textbf{45} (1990), 77--94.

\bibitem{mitchell1993}
Melanie Mitchell, Peter~T. Hraber, and James~P. Crutchfield, \emph{Revisiting
  the edge of chaos: Evolving cellular automata to perform computations},
  Complex Systems \textbf{7} (1993), 89--130.

\bibitem{papadopoulos2024}
Vassilis Papadopoulos, Guilhem Doat, Arthur Renard, and Cl\'{e}ment Hongler,
  \emph{Looking for complexity at phase boundaries in continuous cellular
  automata}, Proceedings of the Genetic and Evolutionary Computation Conference
  Companion, 2024.

\bibitem{papadopoulos2025}
Vassilis Papadopoulos and Etienne Guichard, \emph{{MaCE}: General mass
  conserving dynamics for cellular automata}, arXiv preprint arXiv:2507.12306
  (2025).

\bibitem{plantec2025}
Erwan Plantec, Gautier Hamon, et~al., \emph{{Flow-Lenia}: Emergent evolutionary
  dynamics in mass conservative continuous cellular automata}, Artificial Life
  (2025).

\bibitem{reggia1993}
James~A. Reggia, Steven~L. Armentrout, Hui-Hsien Chou, and Yun Peng,
  \emph{Simple systems that exhibit self-directed replication}, Science
  \textbf{259} (1993), 1282--1287.

\bibitem{rosas2019}
Fernando~E. Rosas, Pedro A.~M. Mediano, Michael Gastpar, and Henrik~J. Jensen,
  \emph{Quantifying high-order interdependencies via multivariate extensions of
  the mutual information}, Physical Review E \textbf{100} (2019), 032305.

\bibitem{sakai2004}
Sunao Sakai, Megumi Kanno, and Yukari Saito, \emph{Phase diagram on a
  generalized cellular automata rule space}, Physical Review E \textbf{69}
  (2004), 066117.

\bibitem{turney2020}
Peter~D. Turney, \emph{Measuring behavioural similarity of cellular automata},
  Artificial Life \textbf{27} (2021), 62--71.

\bibitem{vonneumann1966}
John von Neumann, \emph{Theory of self-reproducing automata}, University of
  Illinois Press, 1966, Edited and completed by Arthur W. Burks.

\bibitem{wolnik2019}
Barbara Wolnik and Bernard De~Baets, \emph{All binary number-conserving
  cellular automata based on adjacent cells are intrinsically one-dimensional},
  Physical Review E \textbf{100} (2019), 022126.

\bibitem{wootters1990}
William~K. Wootters and Christopher~G. Langton, \emph{Is there a sharp phase
  transition for deterministic cellular automata?}, Physica D: Nonlinear
  Phenomena \textbf{45} (1990), 95--104.

\bibitem{yang2024}
Bo~Yang, \emph{Emergence of self-replicating hierarchical structures in a
  binary cellular automaton}, Artificial Life \textbf{31} (2025), 96--105.

\end{thebibliography}

\section*{Acknowledgements}
The author thanks the Cambridge Trust and the Doctoral Training Programme in Medical Research for their support.

\section*{Ethics}
This work is computational and involves no human participants, animal subjects, or personal data. No ethics approval was required.

\section*{AI and AI-assisted technologies}
GitHub Copilot was used for writing assistance and grammar checking. All outputs were reviewed, verified, and edited by the author, who takes full responsibility for the content.

\section*{Data accessibility}
All code and data are available at \url{https://github.com/Don-Yin/self-replication}. The pipeline is fully reproducible from the single entry point \texttt{run.py}. Intermediate results (JSON) and publication figures (PNG) are included in the repository.

\section*{Declaration of AI use}
See AI and AI-assisted technologies statement above.

\section*{Competing interests}
The author declares no competing interests.

\section*{Authors' contributions}
D.Y.: Conceptualization, Methodology, Software, Formal analysis, Investigation, Data curation, Writing (original draft), Writing (review and editing), Visualization.

\section*{Funding}
D.Y. is supported by the Doctoral Training Programme in Medical Research (DTP-MR), University of Cambridge School of Clinical Medicine, and by the Cambridge Trust.

\end{Backmatter}


\printglossary[type=\acronymtype, title=Acronyms and Terms]


\end{document}